# Electronic Correlation-driven Exotic Quantum Phase Transitions in Infinite-layer Manganese Oxide

Heng Jin[1,2] and Bing Huang[2,1, *]

[1]*Department of Physics, Beijing Normal University, Beijing 100875, China*
[2]*Beijing Computational Science Research Center, Beijing 100193, China*

\* Corresponding author: bing.huang@csrc.ac.cn

**Despite the intensive interest in copper- and nickel-based superconductivity in infinite-layer structures, the physical properties of many other infinite-layer transition-metal oxides remain largely unknown. Here we unveil, by the first-principles calculations, the electronic correlation-driven quantum phase transitions in infinite-layer $SrMnO_2$, where spin and charge orders are strongly interwoven. At weak electronic correlation region, $SrMnO_2$ is a ferromagnetic metal with anisotropic spin transportation, as a promising spin valve under room-temperature. At middle electronic correlation region, a structural transition accompanied by charge/bond disproportion occurs as a consequence of Fermi surface nesting, resulting in a ferromagnetic insulator with reduced Curie temperature. At strong electronic correlation region, another structural transition occurs that drives the system into degenerately antiferromagnetic insulators with tunable magnetic order by piezoelectricity, a new type of multiferroics. Therefore, infinite-layer $SrMnO_2$ is possibly a unique system on the quantum critical point, where electronic correlation can induce noticeable Fermi surface evolutions and small perturbations can realize remarkable quantum phase transitions.**

*Introduction.* Many fascinating phenomena in condensed matter physics, including magnetism, ferroelectricity, colossal magnetoresistance, quantum Hall effect, superconductivity, charge/orbital/spin order [1-8], are observed in transition metal oxides (TMOs) with a typical formula of $ABO_3$. With decades of efforts, the synthesis of TMOs becomes mature and efficient [9,10], which provides new opportunities to generate artificial oxides by chemical doping, alloying or topotactic reduction. Special interest of reduction reaction arises from the discovery of copper-based superconductivity in infinite-layer phase [11,12], and is accelerated by newly-discovered nickel-based superconductivity with a similar structure [13]. In general, an infinite-layer TMO with a formula of $ABO_2$ consists of quasi-2D $BO_2$ plane separated by A-dominated charge-reservoir spacer layer, whose physical properties are strongly influenced by different crystal-field splitting and large shift of Fermi level due to the escape of apical oxygen from $ABO_3$. Accompanied by the synthesis and characterization of infinite-layer $ABO_2$ with B=Cu (e.g., $CaCuO_2$ [11,12]), Ni (e.g., $NdNiO_2$ [13]), Fe (e.g., $SrFeO_2$ [14,15]), and Co (e.g., $CaCoO_2$ [16]), an interesting question arises: as the neighbor of existing $ABO_2$ in the element periodic table, can $AMnO_2$ exhibit new physics different from the existing $ABO_2$?

Electronic correlation (EC) is indispensable in understanding charge/spin order, non-Fermi liquid behavior, Mott insulator and unconventional superconductivity [17,18]. The EC is usually strong in systems containing open *d* or *f* shell because of narrow bandwidth and highly-degenerate atomic orbitals. The simplest model to describe the EC effect is the Hubbard model, $H = -\sum_{\langle i,j \rangle} t_{ij}(c_{i\sigma}^\dagger c_{j\sigma} + c_{j\sigma}^\dagger c_{i\sigma}) + U\sum_i n_{i\uparrow}n_{i\downarrow}$, where *i* and *j* are sites, $\sigma$ denotes spin and *U* represents the strength of EC, giving successful interpretation on the correlation-driven Mott phase transition. Density functional theory (DFT) favors of this model [19-21] could give good description on the ground-state structures and electronic structures on the mean-field level as precursor for many-body phenomena. For example, antiferromagnetic (AFM) order in $SrFeO_2$ [22], geometric frustration in $CaCoO_2$ [16], and self-doping effect in $NdNiO_2$ [23] can be well understood by the DFT+*U* calculations. Remarkably, the mentioned physical properties of $ABO_2$ (B=Fe, Co and Ni) are insensitive to the small change of *U*.

In this Letter, we unveil, by the DFT+*U* calculations (see Methods [24]), the electronic correlation-driven exotic quantum phase transitions in infinite-layer $SrMnO_2$, where spin and charge orders are strongly interwoven. At small *U* region, $SrMnO_2$ is a ferromagnetic metal with anisotropic spin transportation. At middle *U* region, a structural transition accompanied with charge/bond disproportion occurs due to Fermi surface nesting (FSN), resulting in a ferromagnetic insulator. At large *U* region, a new structural transition occurs that can transform the system into degenerately AFM insulators with tunable magnetic order by piezoelectric response, resulting in a new type of multiferroics. Our study demonstrates that infinite-layer $SrMnO_2$ is possibly a dreamed system with quantum criticality, differing from other known $ABO_2$ systems.

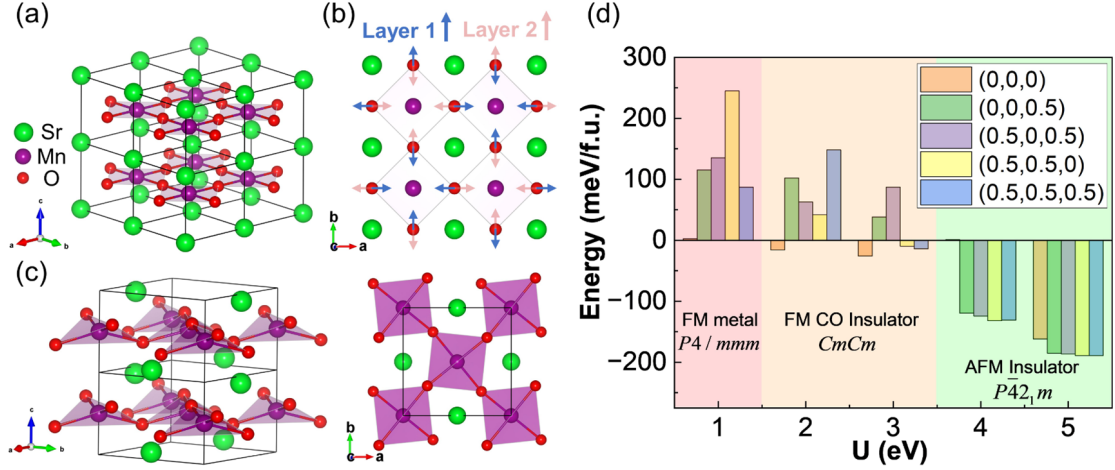

**Figure 1. Structural evolution of SrMnO$_2$ with $U$.** (a) Infinite-layer SrMnO$_2$ with $P4/mmm$ symmetry. Mn-O plane is illustrated with shadowed squares. (b) Schematic diagram of frozen breathing mode in charge/bond disproportion state of SrMnO$_2$ with $CmCm$ symmetry. Arrows with different color indicate movement of oxygens in different Mn-O layers (denoted as Layer 1 and Layer 2). (c) Side view (left) and top view (right) of SrMnO$_2$ with $P\bar{4}2_1m$ symmetry. (d) Total energy versus $U$ of geometrically relaxed SrMnO$_2$ within a $2\times2\times2$ supercell under different magnetic configurations. Energy of FM $P4/mmm$ phase is set to zero. CO, charge order.

*Structural transition induced by EC.* In the topotactic reduction process, the apical O are removed from the perovskite SrMnO$_3$ structure, accompanied by the reduction of space group from $Pm\bar{3}m$ to $P4/mmm$. As shown in **Fig. 1(a)**, the SrMnO$_2$ structure is characterized by stacking Mn-O plane and interstitial Sr atom along $c$ axis. We have evaluated the ground-state transition as a function of $U$. Specifically, one ferromagnetic (FM) [(0,0,0)] and four AFM [(0,0,0.5), (0.5,0,0.5), (0.5,0.5,0) and (0.5,0.5,0.5)] orders are considered (**Fig. S1** [24]). As shown in **Fig. 1(d)**, the $P4/mmm$ phase is only stable with small $U$ (~1 eV), being a FM metal. Unexpectedly, when $U$ is increased to 2~3eV, additional movements of O atoms can be found [**Fig. 1(b)**] to reach a different ground-state configuration. Four O atoms move toward central Mn and away from nearest-neighbor Mn within Mn-O plane, forming frozen breathing mode in Layer-1 and antiphase modulation in neighboring Mn-O plane (Layer-2). Therefore, the structure is reconstructed with $CmCm$ space group and contains $2\times2\times2$ formula unit (f.u.) in the primitive cell, with obvious charge/bond disproportion order. When $U$ is further increased to 4~5eV, the ground-state configuration of SrMnO$_2$ is featured by bent Mn-O plane, rotation of Mn-O tetrahedrons and displacement of Sr along $c$ axis [**Fig. 1(c)**]. This structure has a space group of $P\bar{4}2_1m$, including $\sqrt{2}\times\sqrt{2}\times2$ f.u. in the primitive cell. Interestingly, the two lowest-energy (0.5,0.5,0) and (0.5,0.5,0.5) AFM configurations are nearly degenerated with an energy difference <1 meV/f.u., despite both are insulator.

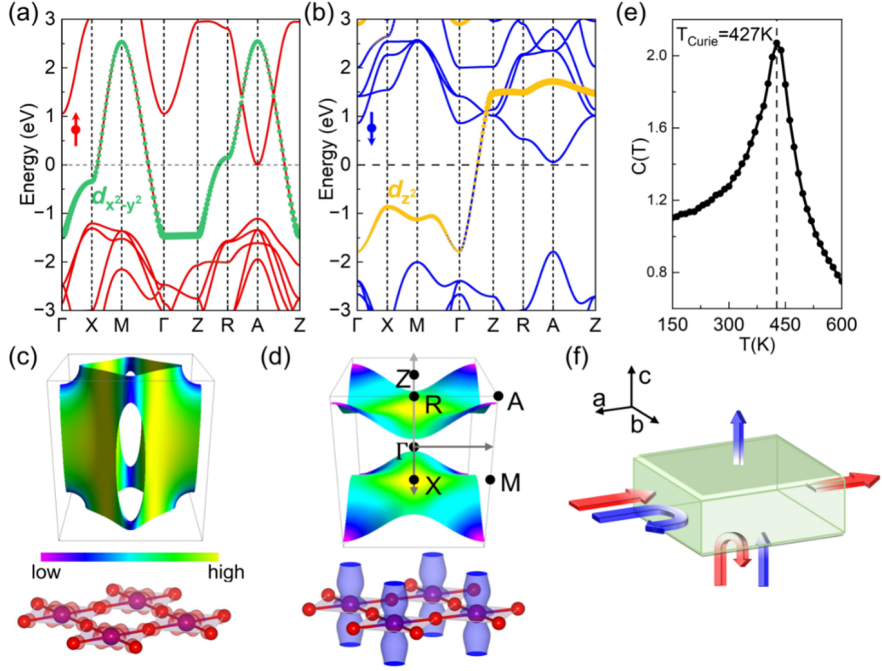

**Figure 2. Electronic properties of ferromagnetic $P4/mmm$ phase with $U=1$eV.** Band structures of (a) spin ↑ and (b) spin ↓ channels. Orbital projections are illustrated with color dots. Fermi level ($E_F$) is set to zero. Fermi surface in Brillioun zone (up pannel) and spin-resovled electron density distribution summed over $E_F \pm 0.1$ eV (down panel) for (c) spin ↑ and (d) spin ↓ band structures. Fermi velocity is illustrated with color bar and electron-density isosurface is set to $2\times10^{-3}$ e/Å$^3$. (e) MC-simulated specific heat $C(T)$ versus temperature based on Heisenberg model. (f) Diagram of anisotropic spin valve. Spin ↑ and spin ↓ currents are illustrated with red and blue, respectively.

*Anisotropic spin valve under small U*. When $U=1$ eV, the ground-state of SrMnO$_2$ is a FM metal. The band structures of both spin channels are shown in **Figs. 2(a)** and **(b)**. The band splitting due to Stoner instability is ~4.25 eV, creating dramatically different electronic states in different spin channels around $E_F$. In spin ↑ channel, $d_{x^2-y^2}$ contributes mainly to the band across $E_F$. Thus, the Fermi surface behaves as a strong quasi-2D feature and the electron density around $E_F$ shows clear in-plane distribution with high Fermi velocity in the 2D Mn-O plane [**Fig. 2(c)**]. Overall, band structure of spin ↑ channel highly resembles infinite-layer nickalates and cuprates [25]. Remarkably, the unoccupied band close to $E_F$ at $A$ is attributed to intersitital electron (**Fig. S2** [24]), which is occupied in NdNiO$_2$ and influences superconductivity fundamentally [26]. The band structure of spin ↓ channel shows a very different feature [**Fig. 2(b)**]. All Mn $d$ orbitals are all pushed above $E_F$ except halfly occupied $d_{z^2}$. The sharp band dispersion along Γ-Z crossing $E_F$ is observed, enabling the Fermi surface of spin ↓ channel to be quasi-1D with high Fermi velocity along Γ-Z [**Fig. 2(d)**]. Therefore, the electron density around $E_F$ shows conducting behavior along the $c$ direction.

From Monte-Carlo (MC) simulation (see Methods [24]), the Curie temperature $T_{Curie}$ estimated by the peak of specific heat is predicted to be ~427 K [**Fig. 2(e)**], far above room-temperature. These

page 4

unique features of Fermi surface of SrMnO$_2$ are highly desirable for designing a spin valve. Common FM metal is conductive for transportation of both spin currents from any direction while half-metal allows transportation of one spin current and prohibits transportation of opposite spin from any direction [27]. In SrMnO$_2$, the transport of spin current is anisotropic. As shown in **Fig. 2(f)**, the transport behavior depends on the direction of spin flows. If electrons are injected along the *a-b* plane, spin ↓ electrons will be filtered out; if the electrons are injected along *c* direction, spin ↑ electrons will be filtered out. Thus, the spin-polarized current can be obtained merely by changing the direction of injected electrons.

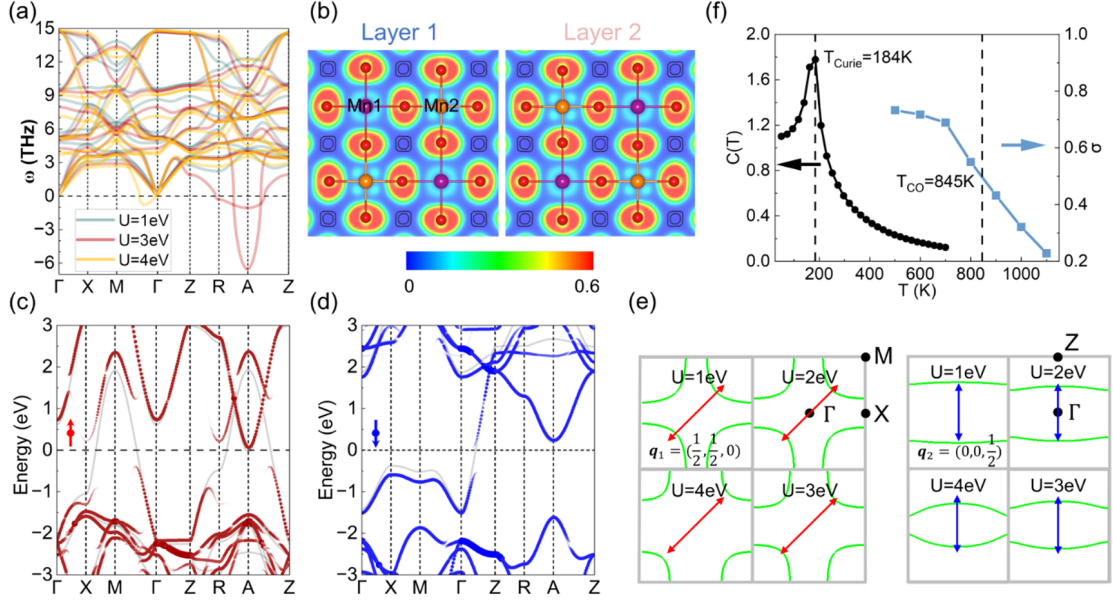

**Figure 3. Electronic properties of charge/bond disproportion phase with *U*=3eV.** (a) Phonon spectrum of *P*4/*mmm* phase under different *U* within FM configuration. (b) Electron localiztion funciton (ELF) of the *CmCm* phase. Two kinds of Mn with different chemical enviroments are colored purple and orange, respectively. Number of layers corresponds to Fig. 1(b). (c) Spin ↑ and (d) spin ↓ unfolded band structures of *CmCm* phase. Band structures of *P*4/*mmm* phase are also plotted with gray lines for comparation. (e) Fermi surface sections across Γ point of *P*4/*mmm* phase under different *U*. Left-panel: spin ↑ with *ab* direction. Right-panel: spin ↓ with *c* direction. Proper nesting vector is decomposed to two vectors, which are labeled in the diagram with red and blue arrows. (f) Specific heat *C*(*T*) and charge-order parameter $\sigma$ versus tempeature from MC simulation and AIMD, respectively.

*Emergent charge/bond disproportion order under middle U.* As shown in **Fig. 1(d)**, when *U* is increased to 2~3 eV, the ground-state is accompanied with charge/bond disproportion. This can be indicated by the phonon spectra of FM *P*4/*mmm* phase [**Fig. 3(a)**], which show darmatic instabilities only when *U*=3 eV. For these two unstable modes, the major imagninary mode reaching -6i THz at *A* is indeed a breathing mode involving the vibiration of Mn-O bonds, as depicted in **Fig. 1(b)**. In the charge/bond disproportion order phase [**Fig. 3(b)**], two kinds of Mn-O bonds are obtained (2.09

Å and 1.95 Å), resulting in two kinds of Mn atoms ($\mu_{Mn1} = 4.3$ $\mu_B$ and $\mu_{Mn2} = 3.4$ $\mu_B$). To unveil the change of electronic states as a consequence of the formation of charge order, the unfolded band structures are plotted and compared with undistorted ones [**Figs. 3(c) and 3(d)**]. Overall, the formation of charge order opens energy gaps in both spin channels, resulting in a FM insulating phase. Specifically, the local gap of spin ↑ channel is about ~1 eV while in spin ↓ channel only a gap ~0.5 eV along Γ-Z can be found.

A curious question is about the driving force behind this charge order. First of all, the Jahn-Teller instability is excluded since no partially-filled degenerate orbitals are found in either spin channel—the degenerate $d_{yz}$ and $d_{xz}$ orbitals of $P4/mmm$ phase are fully occupied (empty) in spin ↑ (↓) channel despite different value of $U$ (**Fig. S3** [24]). Instead, we suggest that FSN may play an important role in driving this charge order. In 1D case, Peierls transition as a consequence of electronic instability and Kohn anomaly result in simultaneous bond disproportion state and gap-opening at $E_F$. Although the power of FSN in forming charge order decreases with the increasing dimensions [28], SrMnO$_2$ could be approximately decoupled to quasi-2D spin ↑ channel and quasi-1D spin ↓ channel, and each part satisfying corresponding decomposed nesting vector. We illustrate this in two steps. First, we verify that the system can be well decoupled. As shown in **Fig. S4** [24], we artificially fixed the vibration of Mn-O bonds in the neighboring Mn-O plane with same phase (antiphase in the $CmCm$ phase). Only the density of states (DOS) of spin ↑ channel around $E_F$ is strongly in reduced, indicating that the gap-opening of spin ↑ (↓) channel is mainly induced by in-plane (out-of-plane) breathing (antiphase) mode. Second, we demonstrate that the nesting condition can be satisfied with decoupled nesting vectors. As shown in **Fig. 3(e)**, the hole pocket of Fermi surface in spin ↑ channel shrinks while the Fermi surface of spin ↓ channel becomes closer to Γ with increasing $U$. The nesting vector forming such $2 \times 2 \times 2$ charge order is $q_{nesting} = (0.5, 0.5, 0.5)$ reciprocal lattice unit (r.l.u.), which can be decomposed as $q_1 = (0.5, 0.5, 0)$ corresponding to spin ↑ channel and $q_2 = (0, 0, 0.5)$ corresponding to the spin ↓ channel. Overall, the nesting condition can be better satisfied when $U=2$ and 3 eV than $U=1$ and 4 eV. Besides, the nesting scenario also agrees with facts that the collapse in phonon spectra is sharp rather than extended and that the major change of both spin channels is near $E_F$ rather than in deep valence band [29].

The application of SrMnO$_2$, such as a platform to study anomalous Hall effect, requires the existence of both charge and FM orders. Thus, it's important to determine the critical temperature of both orders. We performed classic MC simulation on spin Heisenberg model to obtain Curie temperature $T_{Curie}$ and ab initio molecular dynamics (AIMD) to obtain the critical temperature of charge order $T_{CO}$. Here, $T_{CO}$ is approximately determined by $\sigma = 0.5$, where $\sigma(T) = D(T)/D(0)$ and $D(T) = |l_1(T) - l_2(T)|/[l_1(T) + l_2(T)]$ with $l_1(T)$ and $l_2(T)$ are mean Mn-O bond length at temperature $T$. As shown in **Fig. 3(f)**, while the calculated $T_{CO}$ is ~ 845 K, the $T_{Curie}$ is estimated to 184 K, lower than the phase without charge order. Remarkably, during the AIMD simulation FM configuration is set artificially, which disappears above $T_{Curie}$. But the charge order also exists in the ground-state dispite magnetic orders based on the geometric relaxation. Therefore, it is convincible that $T_{Curie}$ ~184 K is the upper-limit temperature for the coexistence of FM and charge-order state.

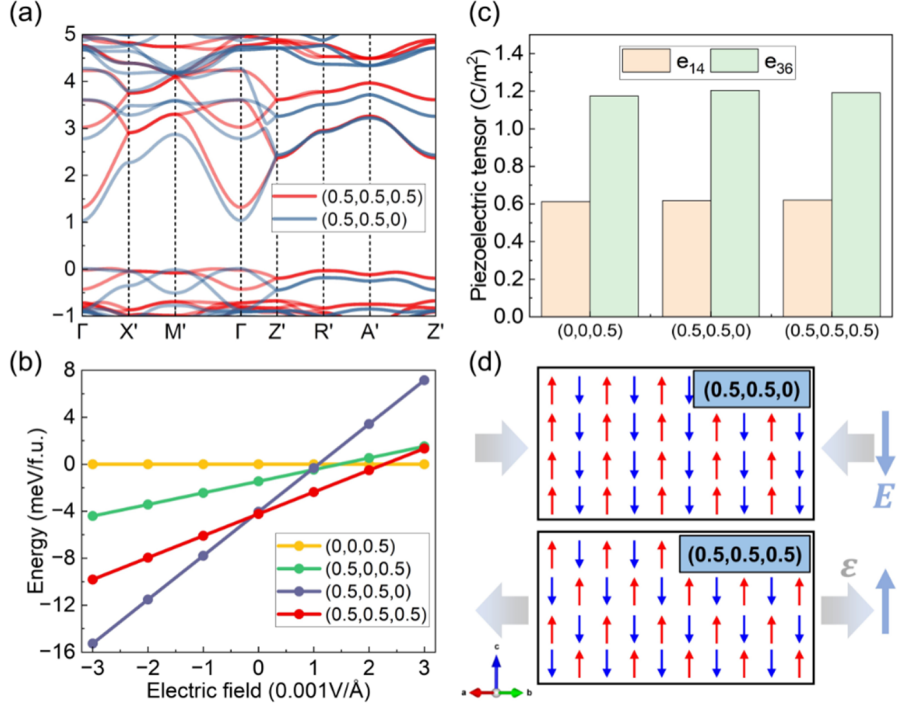

**Figure 4. Electronic properties of nearly-degenerate AFM insulating phases with $U$=5eV.** (a) Band structures of $P\bar{4}2_1m$ phase with (0.5,0.5,0) and (0.5,0.5,0.5) configurations. (b) Energy of AFM configurations under external electric field. Here the electric filed is applied to $c$ axis. (c) Piezoelectric coefficients of $P\bar{4}2_1m$ phase with (0,0,0.5), (0.5,0.5,0) and (0.5,0.5,0.5) configurations. (d) Schematic diagram of tunable antiferromagnetism by piezoelectricity. Red and blue arrows represent different spin orientation. $\varepsilon$ is external strain filed and $E$ is piezoelectricity-induced electric field.

*Piezoelectricity-switchable AFM order under large U.* SrMnO$_2$ is predicted to be AFM metal with $U$=5.3 eV in a previous DFT calcualtion without considering dynamic stabilities [30]. By contrast, in this study, when $U$=5 eV, despite different magnetic configurations large strucuctual reconstructions occur, eliminating the charge order and producing nearly-degenerate (0.5,0.5,0)/(0.5,0.5,0.5) AFM ground states. As shown in **Fig. 4(a)**, both configurations are insulator with a nearly direct bandgap of ~1 eV at Γ point. However, there is distinct difference between the electronic structures of these two configurations. For example, the valence-band dispersion of (0.5,0.5,0) configuration around $E_F$ is larger than that of (0.5,0.5,0.5) configuration. As a consequence, a different response to external electric field ($E_{ext}$) is expected. **Fig. 4(b)** shows the energy of different AFM configurations under $E_{ext}$ parallel to $c$ axis. This suggests (0.5,0.5,0) configuration be the ground state when $E_{ext}$ < 0 V/Å, (0.5,0.5,0.5) configuration be the ground state when 0 < $E$ < 0.002 V/Å, and (0,0,0.5) configuration be the ground state when $E$ > 0.003 V/Å. The magnitude of $E_{ext}$ needed for the magnetic phase transition is quite small (in the order of 10 MV/m) and can be achieved experimentally [31].

On the other hand, emerging piezoelectricity is observed in $P\bar{4}2_1m$ SrMnO$_2$ as a result of non-centric symmetry, which is absent in $P4/mmm$ (small $U$ case) and $CmCm$ (middle $U$ case) phases. The piezoelectric coefficient tensor has two nonzero independent components $e_{14}=e_{25}$ and $e_{36}$. Unlike a $e_{33}$ which reflects the polarization change along $c$ due to strain along the same axis (e.g., in PbZr$_x$Ti$_{1-x}$O$_3$), $e_{36}$ reflects the polarization change along $c$ with response to in-plane strain. As shown in **Fig. 4(c)**, the difference of piezoelectric coefficients of all three AFM configurations is minor ($e_{14}$ ~0.6 C/m$^2$ and $e_{36}$ ~1.2 C/m$^2$), in agreement with similar structures for different magnetic orders. Besides, since electric field along $c$ emerges with in-plane strain as a consequence of nonzero $e_{36}$, the piezoelectric effect may also switch the magnetic ground state. As shown in **Fig. 4(d)**, when a compressive in-plane strain is applied, an electric field along -$c$ is generated, favoring (0.5,0.5,0) configuration; and when a tensile in-plane strain is applied, the direction of electric field is reversed, favoring (0.5,0.5,0.5) configuration. This interesting phenomenon may provide new idea in designing device with AFM order switching by piezoelectricity, a new type of multiferroics.

*Outlook and summary.* The uniqueness of SrMnO$_2$ is observed compared with other infinite-layer systems. While small change of $U$ (of several eV) doesn't influence the basic properties of other ABO$_2$ qualitatively (SrFeO$_2$ (3-6 eV) [22], CaCoO$_2$ (2-6 eV) [16] and NdNiO$_2$ (3-6 eV) [23]), the physical properties of ground-state SrMnO$_2$ is dramatically changed under different $U$. These properties are highly phase-dependent and their existence is associated with the dynamical stability of phase. Basically, we see the electronic states around $E_F$ seem to play an important role in determining the dynamical stability. Varying $U$ by several eV would change the FM band splitting dramatically, accompanied by consequentially changed Fermi surface and Mn-O bonding states of both spin channels. On the other hand, the sensitivity to $U$ indicates that SrMnO$_2$ is indeed on the quantum critical point. Thus, remarkable phase transition can be expected even with small stimulation. Noticeably, in TMOs $U$ can be dependent on local chemical environment or carrier doping. For example, in PrNiO$_2$ [32] $U$ is ~5 eV while in LaNiO$_2$ [33] $U$ is ~6 eV by experimental measurements. Once the synthesis of SrMnO$_2$ is realized, the tunability between different phases can be promising by means of chemical doping or alloying.

In conclusion, this study unveils very unexpected phase transitions in SrMnO$_2$, which is sensitive to the small change of EC. Abundant structural transitions coupled with charge and spin degrees of freedom with novel quantum phenomena are predicted. In the low EC range the system is predicted to be ferromagnetic metal with promising application in spin valve device; with middle EC ferromagnetic charge/bond disproportion order occurs related to correlation-driven FSN; When EC is large, obvious structural reconstruction into $P\bar{4}2_1m$ symmetry with emergent nearly-degenerate

AFM orders is predicted, which can be tuned by electric field induced by piezoelectricity effect, achieving coupled antiferromagnetic order with polarization. Besides, AMnO$_3$, as a star material in colossal magnetoresistance effect [34], has played important roles in information storage. Our study would stimulate experimental interest on SrMnO$_2$ and to check whether its unique properties can be as fascinating as AMnO$_3$.

*Acknowledgements.* This work is supported by the National Key Research and Development of China (Grant No. 2022YFA1402400), NSFC (Grant No. 12088101) and NSAF (Grant No. U2230402). Computations are done at Tianhe-JK supercomputer at CSRC.

**References.**
[1]  P. Zubko, S. Gariglio, M. Gabay, P. Ghosez, and J.-M. Triscone, Annu. Rev. Condens. Matter Phys. **2**, 141 (2011).
[2]  D. Xiao, W. Zhu, Y. Ran, N. Nagaosa, and S. Okamoto, Nat. Commun. **2**, 596 (2011).
[3]  Y. Tokura, M. Kawasaki, and N. Nagaosa, Nat. Phys. **13**, 1056 (2017).
[4]  B. Keimer and J. E. Moore, Nat. Phys. **13**, 1045 (2017).
[5]  R. Ramesh and D. G. Schlom, Nat. Rev. Mater. **4**, 257 (2019).
[6]  S. Dong, H. Xiang, and E. Dagotto, Natl. Sci. Rev. **6**, 629 (2019).
[7]  C. Ahn, A. Cavalleri, A. Georges, S. Ismail-Beigi, A. J. Millis, and J.-M. Triscone, Nat. Mater. **20**, 1462 (2021).
[8]  F. Trier, P. Noël, J.-V. Kim, J.-P. Attan, L. Vila, and M. Bibes, Nat. Rev. Mater. **7**, 258 (2021).
[9]  W. S. Choi *et al.*, Adv. Mater. **24**, 6423 (2012).
[10] Q. Lei *et al.*, npj Quantum Mater. **2**, 10 (2017).
[11] M. G. Smith, A. Manthiram, J. Zhou, J. B. Goodenough, and J. T. Markert, Nature **351**, 549 (1991).
[12] M. Azuma, Z. Hiroi, M. Takano, Y. Bando, and Y. Takeda, Nature **356**, 775 (1992).
[13] D. Li, K. Lee, B. Y. Wang, M. Osada, S. Crossley, H. R. Lee, Y. Cui, Y. Hikita, and H. Y. Hwang, Nature **572**, 624 (2019).
[14] Y. Tsujimoto *et al.*, Nature **450**, 1062 (2007).
[15] C. Tassel *et al.*, J. Am. Chem. Soc. **130**, 3764 (2008).
[16] W. J. Kim *et al.*, Nature **615**, 237 (2023).
[17] B. Keimer, S. A. Kivelson, M. R. Norman, S. Uchida, and J. Zaanen, Nature **518**, 179 (2015).
[18] S. Paschen and Q. Si, Nat. Rev. Phys. **3**, 9 (2020).
[19] V. I. Anisimov, J. Zaanen, and O. K. Andersen, Phys. Rev. B **44**, 943 (1991).
[20] V. I. Anisimov, F. Aryasetiawan, and A. I. Lichtenstein, J. Phys. Condens. Matter **9**, 767 (1997).
[21] S. L. Dudarev, G. A. Botton, S. Y. Savrasov, C. J. Humphreys, and A. P. Sutton, Phys. Rev. B **57**, 1505 (1998).
[22] H. J. Xiang, S.-H. Wei, and M.-H. Whangbo, Phys. Rev. Lett. **100**, 167207 (2008).
[23] E. Been, W.-S. Lee, H. Y. Hwang, Y. Cui, J. Zaanen, T. Devereaux, B. Moritz, and C. Jia, Phys. Rev. X **11**, 011050 (2021).
[24] see Supplemental Material that includes calculation method, spin exchange parameters, magnetic configurations and electronic structure of P4/mmm phase and of artificially fixed phase, and Ref. [35-47].
[25] A. S. Botana and M. R. Norman, Phys. Rev. X **10**, 011024 (2020).

[26] Y. Nomura and R. Arita, Rep. Progr. Phys. **85**, 052501 (2022).

[27] S. D. Bader and S. S. P. Parkin, Annu. Rev. Condens. Matter Phys. **1**, 71 (2010).

[28] M. D. Johannes and I. I. Mazin, Phys. Rev. B **77**, 165135 (2008).

[29] K. Rossnagel, J. Phys. Condens. Matter **23**, 213001 (2011).

[30] M. Rahman, K.-C. Zhou, Y.-Z. Nie, and G.-H. Guo, Solid State Commun. **266**, 6 (2017).

[31] T. Lottermoser, T. Lonkai, U. Amann, D. Hohlwein, J. Ihringer, and M. Fiebig, Nature **430**, 541 (2004).

[32] Z. Chen *et al.*, Matter **5**, 1806 (2022).

[33] M. Hepting *et al.*, Nat. Mater. **19**, 381 (2020).

[34] A. P. Ramirez, J. Phys. Condens. Matter **9**, 8171 (1997).

[35] S. Nos, J. Chem. Phys. **81**, 511 (1984).

[36] W. G. Hoover, Phys. Rev. A **31**, 1695 (1985).

[37] P. E. Blöchl, Phys. Rev. B **50**, 17953 (1994).

[38] G. Kresse and J. Furthmüller, Phys. Rev. B **54**, 11169 (1996).

[39] J. P. Perdew, K. Burke, and M. Ernzerhof, Phys. Rev. Lett. **77**, 3865 (1996).

[40] S. L. Dudarev, G. A. Botton, S. Y. Savrasov, C. J. Humphreys, and A. P. Sutton, Phys. Rev. B **57**, 1505 (1998).

[41] J. H. Lee and K. M. Rabe, Phys. Rev. Lett. **104**, 207204 (2010).

[42] P. V. C. Medeiros, S. Stafström, and J. Björk, Phys. Rev. B **89**, 041407 (2014).

[43] P. V. C. Medeiros, S. S. Tsirkin, S. Stafström, and J. Björk, Phys. Rev. B **91**, 041116 (2015).

[44] A. Togo and I. Tanaka, Scripta Mater. **108**, 1 (2015).

[45] A. Edström and C. Ederer, Phys. Rev. Mater. **2**, 104409 (2018).

[46] M. Kawamura, Comput. Phys. Commun. **239**, 197 (2019).

[47] L. Liu, X. Ren, J. Xie, B. Cheng, W. Liu, T. An, H. Qin, and J. Hu, Appl. Surf. Sci. **480**, 300 (2019).